128

# Combining Unsupervised and Supervised Learning for Asset Class Failure Prediction in Power Systems

Ming Dong, *Senior Member, IEEE*

*Abstract*—In power systems, an asset class is a group of power equipment that has the same function and shares similar electrical and/or mechanical characteristics. Predicting failures for different asset classes is critical for electric utilities towards developing cost-effective asset management strategies. Previously, physical age based Weibull distribution has been widely used for failure prediction. However, this mathematical model cannot incorporate asset condition data. As a result, the prediction cannot be very specific and accurate for individual assets. To solve this important problem, this paper proposes a novel and comprehensive data-driven approach based on asset condition data: K-means clustering as an unsupervised learning method is used to analyze the inner structure of historical asset condition data and produce the asset conditional ages; logistic regression as a supervised learning method takes in both asset physical ages and conditional ages to classify and predict asset operation statuses. Furthermore, an index called average aging rate is defined to quantify, track and estimate the relationship between asset physical age and conditional age. This approach was applied to a medium-voltage cable class in an urban distribution system in West Canada. Case studies and comparison with standard Weibull distribution are provided. The proposed approach demonstrates higher accuracy measured by F1-Score than Weibull distribution method for asset class failure prediction.

## I. Introduction

TODAY, more and more electric utilities are mandated by regulators to develop cost-effective long-term asset management strategies to reduce overall cost while maintaining system reliability [1-2]. Sophisticated and optimal asset management strategies can only be established based on the accurate prediction of asset failures in the future. For example, knowing the number of service transformer failures in the next few years, electric utilities can purchase and stock enough spares and prepare necessary working resources to deal with potential failure events; electric utilities can also proactively replace a certain number of service transformers to reduce the potential failures. In return, the system reliability can be maintained and the asset risks can be minimized.

Previously, some research works applied machine learning and advanced data analytics to predicative maintenance and condition monitoring tasks which mainly target critical assets and rely on the use of special sensors or testing devices [3-8]. In such applications, the health prediction is based on the learning and analysis of measured data against the asset's health status. Two types of historical data can be collected for learning: data from process sensors such as speed sensors or angle sensors that reflect the working status [3]; data from testing sensors or devices such as voltage sensors and oil sample tanks that measure certain auxiliary signals for diagnosis purpose [4-5]. If online sensors are installed, continuous time-series data can also be processed using trending analysis [6]. Sometimes, the predicted asset health status can be attributed to different fault causes and different severity levels and multiclass classification methods can serve in such cases [7-8]. According to the classification result, proper inspection and maintenance decisions can be made to ensure the expected operation of the asset. Essentially, predictive maintenance focuses on the diagnosis of the current health status of a certain critical asset based on the measurement data. It is not suitable for long-term prediction problem such as the prediction in 5 to 10 years.

In comparison, asset class failure prediction addresses a different problem related to long-term asset management strategy. Many power system assets can be grouped as asset classes in which the equipment has the same function and shares similar electrical and/or mechanical characteristics. There are many different asset classes in both transmission and distribution systems. Examples of transmission asset classes are a certain type of transmission towers, transmission overhead conductors, transmission underground cables and etc.; examples of distribution asset classes are a certain type of distribution poles, distribution overhead conductors, distribution underground cables, distribution service transformers and etc. Individually, these assets are typically much cheaper than the aforementioned critical assets and are therefore not equipped with special sensors or testing devices that can support more granular health diagnosis. Due to the large quantity and low value of these assets, it is common for utility companies to consider only two asset operation statues for the long-term management of these assets: working or failed. A failed asset will be replaced. The main purpose of asset class failure prediction is to support the determination of overall availability and reliability risk of an asset class in the

M. Dong is with Department of System Planning and Asset Management, ENMAX Power Corporation, Calgary, AB, Canada, T2G 4S7 (e-mail: mingdong@ieee.org).

future. Based on this, utility companies can further plan for proactive asset replacement and spare stock level in a scientific way. A unique advantage of predicting asset failures in an asset class is the ability to leverage historical asset data of other assets in the same asset class since the assets in one class are of the same type and therefore generally follow similar aging or degradation processes.

Traditionally, Weibull distribution has been widely used by utility asset engineers for asset class failure prediction [9-16]. Asset failures and asset physical ages at the time of failures are recorded, analyzed and modeled by Weibull distribution functions. Mathematically, the standard cumulative Weibull distribution function is given as:

$$F(A^P) = 1 - e^{-\left(\frac{A^P}{\alpha}\right)^\beta}, \quad A^P \geq 0 \qquad (1)$$

where $\alpha$ is the scale parameter; $\beta$ is the shape parameter; $A^P$ is the asset's physical age; $F(A^P)$ is the cumulative probability of failure at age $A^P$. By analyzing failure history of an asset class, optimal values of $\alpha$ and $\beta$ can be determined so that the above distribution function can statistically represent the failure process of this type of asset. Typically, a degradation curve or a survival curve can be produced in the end. For any hypothetical age point, the corresponding asset failure probability can be obtained from the curve. This way, the number of failures for an asset population in the long-term future can be predicted and the associated risks can be analyzed.

This classic method, however, has its limitations. In reality, physical age is only one attribute of an asset. Individual assets at the same physical age can have significantly different health conditions. This is because individual assets could have been operated under different modes such as different voltage and loading levels; they could also have been maintained in different ways such as different maintenance frequencies. The prediction that solely relies on physical age may be able to generate a reasonable statistical view for an asset population when the size of population is large. But it is not very reliable for predicting individual assets' operation statuses because asset condition data is completely ignored. Fortunately, in recent years, many electric utility companies have realized the value embedded in big data and started to introduce Computerized Maintenance Management Systems (CMMS) along with sophisticated asset inspection/testing programs to gather, track and store asset condition data [12-15]. Different from asset failure data, asset condition data contains much more information specific to individual assets: it can include inspection and testing results which directly reflect the asset's health conditions; the dataset can also include long-term data in which the condition variation information is kept and can be analyzed for prediction purposes.

To address the limitations of Weibull distribution based methods, this paper proposes a novel and comprehensive approach to predict asset class failures for power systems. The contributions of this work include:
- It devises a unique learning process to analyze the inner structure of historical asset condition data and model the statistical relationship between an asset operation status and the asset condition data;
- It devises a unique prediction process to predict the change of asset conditions in the long-term future based on newly proposed indices asset conditional age and average aging rate;
- It proposes the use of sophisticated unsupervised and supervised learning models instead of Weibull distribution function for asset class failure prediction;
- It proposes a model to effectively incorporate the asset condition data. As a result, more accurate and specific prediction of individual asset operation status in an asset class can be produced.

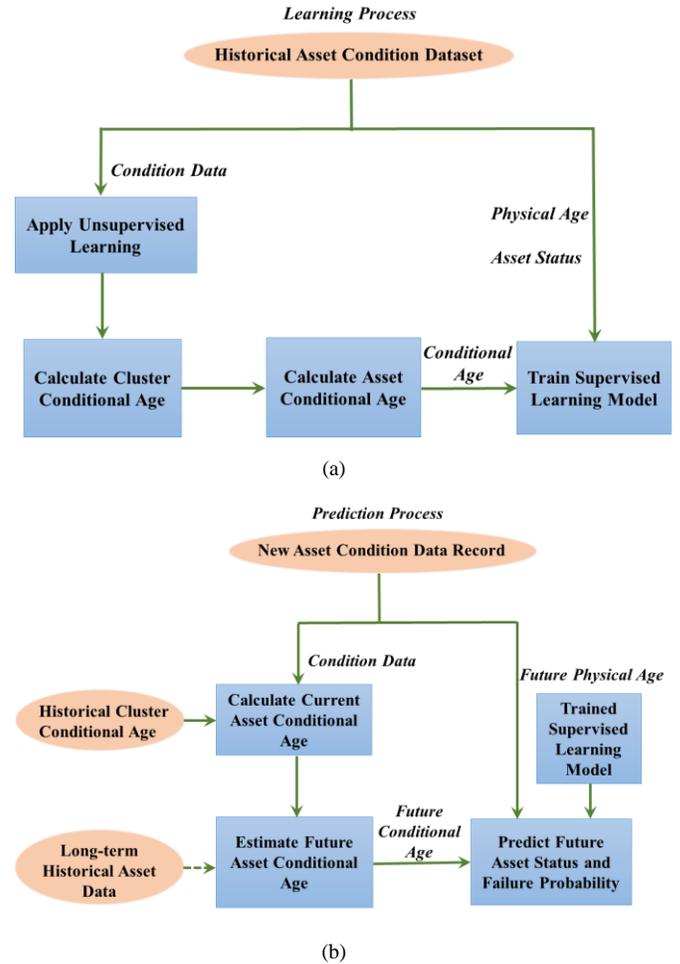

Fig.1. Flowchart of the proposed approach

This new approach contains two processes: the learning process and the prediction process. The learning process is illustrated in Fig.1(a): through K-means clustering based unsupervised learning, historical asset condition data can be automatically grouped into clusters based on internal similarities. The conditional ages of individual clusters and assets are then calculated. Physical ages, conditional ages and asset operation statuses (working or failed) are fed into the final step to train a logistic regression classifier which is used as the supervised learning model. At the end of this learning process, the logistic regression classifier will be trained and can automatically classify an asset into either working or failed status based on its conditional and physical ages.

After the learning process, a new asset condition data record is given to predict the future asset operation status and failure probability. This process is illustrated in Fig.1(b): first, the asset condition data is used to calculate the current asset conditional age. After the step, this paper defines a new index called average aging rate to estimate this asset's future conditional age. If there is long-term historical asset data, the variation of average aging rates of similar assets can be analyzed and used as a reference. This will ensure more accurate estimation of future conditional age. In the end, using both future conditional and physical ages, the asset status can be predicted by the logistic regression classifier which is trained previously during the learning process.

The main body of this paper is organized as below: in the beginning, this paper introduces the definitions and categories of asset condition data; it then explains the learning process which covers the steps of unsupervised learning, conditional age calculation and supervised learning; in Section IV, it explains the prediction process which covers the steps of the estimation of future conditional age and the prediction of asset operation status and failure probability by using the established logistic regression classifier.

In the end, this approach was applied to a medium-voltage cable class in an urban distribution system in Western Canada and multiple case studies are given. The results are compared to the ones produced by a traditional Weibull distribution model. It is found that the proposed approach has better performance for predicting asset class failures in power systems.

## II. ASSET CONDITION DATASET

In recent years, more and more electric utility companies are moving towards condition-based maintenance strategy. This strategy is recommended by ISO 55000 and encouraged by many utility regulators [21]. On the one hand, utility companies have been establishing sophisticated inspection and testing programs to gather asset health condition data [20]; on the other hand, utility companies now track and keep asset condition data using specialized CMMS software systems such as IBM Maximo [17-19]. Before discussing the proposed learning process and prediction process, it is necessary to understand the structures of asset condition data that is used in these processes. In this paper, asset condition data should include three parts: health condition data, physical age and asset operation status. Asset operation status is a binary status, i.e. working or failed at the data recording time. Historical asset condition dataset has known asset operation statuses; future asset operation statues are not known but will be predicted by using the proposed approach. It should be noted that in reality, for a specific asset class, its asset data may not be directly stored by CMMS in the format exemplified below. However, proper data manipulation can be employed to consolidate data records into the desired format.

The health condition data can be acquired through asset inspection and testing. Different health condition features are acquired for different asset classes [20]. In practice, condition data is often organized by asset class. For example residential single-phase pad-mount transformers and residential single-phase overhead transformers are normally considered as two different asset classes. Each of them has its own asset condition dataset. Depending on the maturity of CMMS system adoption, asset data management practice and asset inspection/testing practice in a utility company, for a specific asset class, there could be long-term condition data or only one-time condition data in place. This difference would result in different methods for estimating future conditional age and will be discussed in detail in Section IV.

An example of one-time condition dataset and long-term condition dataset are shown in Table I and II. In these examples, an asset's health condition data has three numerical features ($H_1^n, H_2^n, H_3^n$) and one categorical feature $H_1^c$. It is important to recognize the difference between numerical features and categorical features. This difference needs to be considered specifically in the unsupervised learning process to be discussed in Section III. Comparing Table II with Table I, it can be found that the long-term dataset contains data recorded in different inspection years at a 3-year interval. Health condition change can be observed between different inspection years. In addition to health condition data, each individual asset's physical age and asset status are also included in the datasets.

TABLE I: AN EXAMPLE OF ASSET ONE-TIME CONDITION DATASET

| Asset ID | Health Condition Data | | | | Age (Year) | Asset Operation Status |
|---|---|---|---|---|---|---|
| | $H_1^n$ | $H_2^n$ | $H_3^n$ | $H_1^c$ | | |
| 0001 | 26 | 1.38 | 198 | Medium | 28 | Working |
| 0002 | 37 | 0.78 | 183 | Medium | 35 | Failed |
| 0003 | 36 | 0.60 | 217 | Severe | 21 | Failed |
| 0004 | 46 | 1.51 | 196 | Moderate | 42 | Working |
| 0005 | 12 | 2.44 | 235 | Moderate | 39 | Working |
| … | … | … | … | … | … | … |

TABLE II: AN EXAMPLE OF ASSET LONG-TERM CONDITION DATASET

| Asset ID | Inspection Year | Health Condition Data | | | | Age (Year) | Asset Operation Status |
|---|---|---|---|---|---|---|---|
| | | $H_1^n$ | $H_2^n$ | $H_3^n$ | $H_1^c$ | | |
| 0001 | 2018 | 26 | 1.38 | 198 | Medium | 28 | Working |
| 0001 | 2015 | 20 | 1.43 | 197 | Medium | 25 | Working |
| 0001 | 2012 | 15 | 1.42 | 201 | High | 22 | Working |
| 0002 | 2018 | 37 | 0.78 | 183 | Medium | 35 | Failed |
| 0002 | 2015 | 32 | 1.55 | 183 | Medium | 32 | Working |
| 0002 | 2012 | 22 | 1.69 | 186 | Medium | 29 | Working |
| 0003 | 2018 | 36 | 0.60 | 217 | Severe | 21 | Failed |
| 0003 | 2015 | 30 | 0.89 | 216 | Severe | 18 | Working |
| 0003 | 2012 | 26 | 1.69 | 221 | Medium | 15 | Working |
| … | | … | … | … | … | … | … |

## III. LEARNING PROCESS

This section explains the details of the learning process in the proposed approach. As shown in Fig.1 (a), this process consists of: applying unsupervised learning to historical asset condition data; calculating cluster conditional age; calculating asset conditional age; and training the supervised learning model.

### A. Unsupervised learning of asset condition data

Unsupervised learning is a type of machine learning that learns from existing data that has not been pre-classified or pre-labeled [22]. Unsupervised learning examines the entire dataset, analyzes the commonalities of data points and groups similar data points together. By applying unsupervised learning to asset condition data, we can understand the inner structure of health conditions for a given asset class. This establishes the foundation for further producing asset conditional ages.

K-Means clustering is selected as the unsupervised learning method to process the asset condition dataset shown in Section II. K-Means method is a widely used clustering method for dealing with large datasets with great efficiency and simplicity. It requires only one input parameter $K$ which is the expected number of clusters for data grouping. Depending on the number of condition features and the variance of condition data, $K$ can be selected accordingly. In practice, to optimize the performance, different $K$ values can be tested and clustering quality evaluation method such as Silhouette analysis can be applied to identify the best $K$ value [23]. This part will be discussed later in this section.

The mathematic description of K-Means clustering is stated as below: given a set of observations $(X_1, X_2, ..., X_n)$, where each observation is a $d$-dimensional real vector, K-Means clustering aims to group $n$ observations into $K$ ($\leq n$) clusters $S = \{S_1, S_2, ..., S_K\}$ so as to minimize the within-cluster variances. Formally, the objective is to find:

$$\arg\min \sum_{i=1}^{K} \sum_{x \in S_i} \|x - \mu_i\|^2 = \arg\min \sum_{i=1}^{K} |S_i| \, Var \, S_i \quad (2)$$

where $\mu_i$ is the mean of data points in $S_i$ [22].

The standard steps of K-Means clustering are:

- Step 1: Initialize $K$ centroids randomly within the data domain;
- Step 2: Associate all data points to their nearest centroids based on Euclidean distance. This step will create $K$ data clusters. Each cluster contains the associated data points as its members;
- Step 3: Update the centroid of each cluster using all members in the cluster;
- Step 4: Repeat Step 2 and Step 3 until convergence has been reached.

As shown in Table I and II, the full feature vector of each condition record can be defined as:

$$V = (H_1^n, H_2^n, ..., H_p^n, H_1^c, ... H_q^c) \quad (3)$$

where $p$ is the number of numerical features; $q$ is the number of categorical features. It should be noted that in practical applications, utility engineers do not necessarily have to use the full condition feature vector. Instead, only features with high variance and high relevance to asset failures need to be selected based on certain domain knowledge of the asset. Alternatively, there are algorithms that can be used to mathematically reduce the dimension of the condition feature vector: principal component analysis can transform correlated features into fewer uncorrelated features [24]; Relief algorithm [25] can be used to find the condition features that are only beneficial to the intra-cluster grouping and inter-cluster separation while excluding unbeneficial features for clustering. These steps will be able to enhance the clustering quality.

As (3) shows, the asset condition data often includes both numerical and categorical condition features. The distance of numerical features between two data points can be calculated using standard Euclidean distance as below:

$$d(X, Y) = \sum_{j=1}^{p} (x_j - y_j)^2 \quad (4)$$

For categorical condition features, two different methods can be used to process them. An orderly categorical condition feature value $c$ such as low, medium, high can be converted into a numerical condition feature value $x$ using equation below:

$$x = \frac{c - 1/2}{N}, c = 1, 2 ... N \quad (5)$$

where $N$ is the total number of orderly statuses, $c$ is the order of the status [26].

Another type of categorical conditional feature is unordered. It often indicates the presence of a certain type of health symptom such as the type of pole appearance damage. If this kind of feature needs to be included, the standard Euclidian distance in (4) should be modified as below:

$$\begin{cases} d(X, Y) = \sum_{j=1}^{p} (x_j - y_j)^2 + \sum_{j=p+1}^{p+q} \delta(x_j, y_j) \\ \delta(x_j, y_j) = \begin{cases} 0, (x_j = y_j) \\ 1, (x_j \neq y_j) \end{cases} \end{cases} \quad (6)$$

where $x_1$ to $x_p$ and $y_1$ to $y_p$ are all numerical features and $x_{p+1}$ to $x_{p+q}$ and $y_{p+1}$ to $y_{p+q}$ are unordered categorical features. If two unordered categorical features match, their Euclidian distance is 0 otherwise is 1 [26].

To effectively apply K-means, all numerical features (including the ones converted from orderly categorical features) should be normalized to a fixed range such as [0,1]. This is because the raw condition data has different units and the difference between feature magnitudes can be quite large. There are many ways to normalize data, for example, the classic Min-Max normalization [22]:

$$x_{norm} = \frac{x_{raw} - Min}{Max - Min} \quad (7)$$

where $Max$ is the maximum value observed in feature $j$; $Min$ is the minimum value observed in feature $j$.

In addition to feature normalization, if utility asset engineers have prior knowledge about the importances of certain tests to the asset failure probability, different weighting factors can be assigned to their corresponding condition features in the Euclidean distance formula (6) when applying K-means clustering. The formula would then become:

$$\begin{cases} d(X, Y) = \sum_{j=1}^{p} w_j (x_j - y_j)^2 + \sum_{j=p+1}^{p+q} w_j \delta(x_j, y_j) \\ \delta(x_j, y_j) = \begin{cases} 0, (x_j = y_j) \\ 1, (x_j \neq y_j) \end{cases} \end{cases} \quad (8)$$

where $w_j$ is an empirical weighting factor for feature $j$ based on the feature importance.

As mentioned previously, Silhouette analysis is a clustering quality evaluation method that can be used to determine the proper parameter $K$ from a range of $K$ values [23]. In this analysis, Silhouette coefficient is used as a metric to evaluate clustering quality. For a certain data point $r \in C_r$, its Silhouette coefficient $S_r$ can be mathematically calculated as below:

$$\begin{cases} S_r = \dfrac{b_r - a_r}{max(a_r, b_r)} \\ a_r = \dfrac{1}{|C_r| - 1} \sum_{s \in C_r, r \neq s} d(r, s) \\ b_r = min \dfrac{1}{|C_t|} \sum_{t \in C_t} d(r, t) \end{cases} \quad (9)$$

where $|C_r|$ is the cardinality of set $C_r$, i.e. the number of members in cluster $C_r$; $C_t$ is any other cluster in the dataset; data point $t$ belongs to $C_t$; $d$ is the Euclidean distance between two data points as expressed previously in (4), (6) and (8).

The above equation evaluates both the compactness and separation of clusters: $a_r$ is the average distance of data point $r$ to all other points in the same cluster $C_r$. It describes the intra-cluster compactness. The smaller $a_r$ is, the more compact $C_r$ is; $b_r$ is the smallest average distance of $r$ to all points in every other cluster in this dataset. It describes the inter-cluster separation. The bigger $b_r$ is, the more inter-cluster separation is seen from point $r$; in the end, $S_r$ combines $a_r$ and $b_r$, aiming to maximize intra-cluster compactness and inter-cluster separation. The Silhouette coefficient is normlaized and ranges between −1 and 1.

It should be noted (9) is the calculation for one single data point $r$. To evaluate the clustering quality of the entire dataset, average Silhouette coefficient can be used and is calculated as below:

$$S_{avg} = \frac{1}{m} \sum_{i=1}^{m} S_i \quad (10)$$

where $m$ is total number of data points in this dataset.

The steps of using Silhouette analysis to determine optimal cluster number $K$ are given as below:

- Step 1: Initialize $K$ to a numerical range;
- Step 2: For a given $K$, apply K-means clustering;
- Step 3: Use (9) and (10) to calculate $S_{avg}$;
- Step 4: Repeat steps 2-3 until $S_{avg}$ for every $K$ in the intial range is calculated. Select the $K$ with largest $S_{avg}$ as the optimal cluster number for K-means clustering.

*B. Calculate cluster conditional age*

This paper proposes an important concept called asset conditional age. Different from asset physical age, asset conditional age is the statistical age derived from the asset's health conditions. It indicates the age that certain asset health conditions likely fall under. Every asset has both physical age and conditional age. For example, a 50-year old transformer has very healthy conditions and if only looking at its health conditions, this transformer appears to be 30-year old. In this case, 50 is the transformer's physical age and 30 is the conditional age. The failure probability of an asset is not only affected by the physical age but also by the conditional age. Two assets at the same physical age could be operated and maintained differently and could therefore reveal different health conditions. For example, some distribution wood poles have communication cables or pole-mount transformer tanks on them. The additional weight may cause surface cracks and reduce pole health. Also, depending on where the poles are located along the power line, there are tangent poles carrying straight-line overhead conductors, angle poles carrying turning conductors and dead-end poles carrying conductors to only one side of the poles. These operating differences result in unbalanced bending forces on the poles and also affect their health conditions. Similarly, maintenance differences significantly affect asset's lifespan and failure probability. Still taking distribution wood poles as an example, it is common for utility companies in North America to treat in-service poles with different chemical preservatives.

Furthermore, the change of asset's health conditions can differ from the change of physical age. For example, a lightly loaded, well-maintained transformer may increase its conditional age much slower than its physical age; a heavily loaded, poorly-maintained transformer may change its conditional age much faster than its physical age.

After the previous unsupervised learning step, assets with similar conditions have been grouped into $K$ condition clusters. Based on this new data structure, the conditional age of each cluster can be calculated as:

$$A_j^C = \frac{\sum_{i=1}^{N} A_i^P}{N} \quad (11)$$

where $A_i^P$ is the physical age of asset $i$ in condition cluster $j$; $N$ is the number of asset members in cluster $j$. Each cluster's conditional age is the average physical age of all members in the cluster. Table III shows an example of 10 clusters' conditional ages for an asset class. These cluster conditional ages can be used as baseline values to calculate an individual asset's conditional age with any condition data in the same asset class.

TABLE III: AN EXAMPLE OF CLUSTER CONDITIONAL AGES

| Cluster ID | # of Members | Conditional Age(Year) |
|---|---|---|
| 1 | 89 | 23.4 |
| 2 | 77 | 22.7 |
| 3 | 49 | 65.4 |
| 4 | 63 | 55.8 |
| 5 | 123 | 15.8 |
| 6 | 11 | 72.0 |
| 7 | 136 | 5.6 |
| 8 | 68 | 38.3 |
| 9 | 25 | 44.9 |
| 10 | 19 | 83.5 |

*C. Calculate conditional age of individual asset*

Once cluster conditional ages are calculated as baseline values, the conditional ages of a given asset with condition vector $X$ can be calculated. One simple way is to choose the conditional age of the nearest cluster as $X$'s conditional age. However, this method is not accurate because it does not consider the gravities from other clusters. Alternatively, this paper proposes the following mathematic equation using Euclidean distances between $X$ and the centroids of all clusters:

$$A_X^C = \sum_{j=1}^{k} A_j^C \left( \frac{\dfrac{1}{d(X, C_j)}}{\sum_{j=1}^{k} \dfrac{1}{d(X, C_j)}} \right) \quad (12)$$

where $A_j^C$ is the conditional age of cluster $j$ previously calculated by (11) ; $C_j$ is the feature vector of the centroid in cluster $j$; $X$ is the condition vector of the interested asset.

The meaning of (12) is explained with respect to Fig.2. A condition vector $X$ is given and it has two features. If $X$ is very close to an existing cluster's centroid $C_j$, the Euclidian distance $d(X, C_j)$ will become close to zero and as a result, (12) would be approximated to $A_x^C \approx A_j^C$. This means $X$'s conditional age should also be very close to cluster $C_j$'s conditional age. In a more common case where $X$ is surrounded by all clusters, $A_x^C$ is the average of all clusters' conditional ages weighted by the similarities between their centroids and $X$. This method could provide reasonable estimation as long as $X$ is not far away from all clusters. If $X$ is far away from all clusters, this means $X$ represents a new condition that was not learned from the historical asset condition data. This suggests that the historical asset condition dataset should contain asset records with a wide range of conditions in case of encountering new conditions.

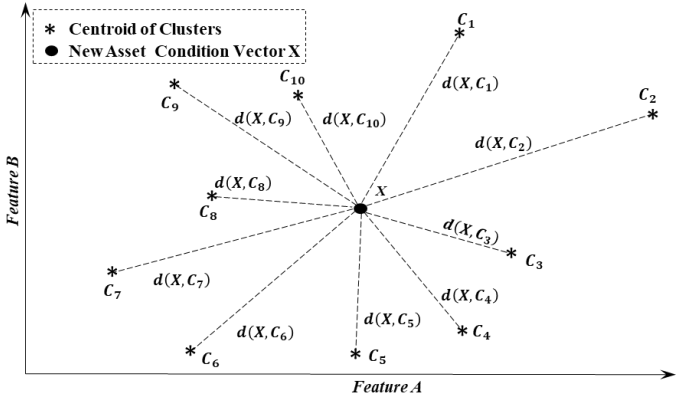

Fig. 2. Calculate cluster conditional age based on condition clusters

### D. Supervised learning of asset condition data

Supervised learning is a type of machine learning that learns from existing data that has been pre-classified or pre-labeled [22]. Classification problem is a typical supervised learning problem. It is expected that after learning, machine will be able to automatically classify objects based on certain input variables. In this research, the goal is to develop a classifier that can classify an asset into either working or failed status based on asset physical age and conditional age. Logistic regression classifier (LRC) has been widely used as a supervised learning model in different disciplines for binary status classification [27-30]. Compared to other binary classifiers such as Support Vector Machine, Neural Network and Decision Tree, LRC is easier to implement and also explicable from the probability perspective. In this research, LRC is chosen as the supervised learning model. Different from Weibull distribution, a multivariate LRC can process more than one input variables, in our case both the physical age and conditional age of an asset. The logistic regression function used in this paper is given as:

$$L = \frac{1}{1+e^{-(\beta_0+\beta_1 A^P+\beta_2 A^C)}} \quad (13)$$

where $\beta_0$, $\beta_1$ and $\beta_2$ are parameters to be determined through model training.

Following standard supervised learning process, the asset condition dataset should be split into a training set and a test set. Training set is used to determine the parameters of the model and the test set is used to evaluate the classification performance of the model.

One important technique that is required during the training of LRC is oversampling of failure records. This is because the numbers of failed asset records and working asset records in the historical asset condition dataset can be very unbalanced. Oftentimes, there are far more working assets than failed assets in the dataset. If directly using the records for training, the trained classifier could become biased towards working status. Oversampling means purposely duplicating the failure records in the training set so that the numbers of failure asset records and working asset records in the training set are approximately the same. This step can ensure a classifier that can only effectively detect the majority (working) status yields a large error. The classifier will become less biased through the training process. As discussed in [30], this technique can effectively improve the training accuracy of binary classifiers.

### IV. PREDICTION PROCESS

This section explains the details of the prediction process in the proposed approach. As shown in Fig.1 (b), this process consists of: calculating current conditional age; estimating future conditional age; and predicting asset future status.

### A. Calculate current conditional age

If a new asset condition data record is given, the asset's conditional age can be calculated following the same process discussed in Section III-C, with respect to the historical cluster conditional ages established during the learning process.

### B. Estimate future conditional age

In order to predict the future asset status and failure probability, the future conditional age has to be estimated first while the physical age can be simply calculated based on the time difference. Directly estimating the change of multiple asset condition values over time can be very challenging because they can vary in different directions and are often correlated with each other. In comparison, estimating the change of asset conditional age as an overall index is much easier due to its clear physical meaning and its natural relationship with the physical age. Similar to physical age, conditional age increases with time. It is obvious that after a physical year, the physical age increment will be one; however the conditional age increment may not be one. Within one physical year, a lightly-loaded, well-maintained transformer may grow its conditional age by only a few months; in contrast, a heavily-loaded, poorly-maintained transformer may deteriorate its health conditions significantly and grow its conditional age by a few years. In order to capture this interesting phenomenon, this paper proposes a new index called average aging rate to describe the relationship between conditional age and physical age. It is defined as:

$$R = \frac{A^C}{A^P} \quad (14)$$

For example, if a service transformer has a 30-year physical age and 60-year conditional age, the average aging rate is 2.

As discussed in Section II, utility companies may or may not have long-term asset condition dataset for a certain asset class. When there is no long-term asset condition dataset, to estimate the future conditional age, it is assumed that the average aging rate will not change significantly in the near future. Therefore the future conditional age can be estimated by:

$$A^C = R \times (A^P + T) \quad (15)$$

where $T$ is the time difference between today and future.

When there is long-term asset condition dataset, a more accurate estimation method becomes available: first, from the historical dataset, a few assets in a historical year that are similar to the target asset are found. This can be done by calculating and ranking the Euclidean distances (including asset conditions and physical age) of the target asset and each historical asset. Those assets with minimal distances will be selected; second, the future aging rate $R^T$ can be calculated by:

$$R^T = R \frac{\sum_{i=1}^{n}\left(\frac{R_i^T}{R_i}\right)}{n} \quad (16)$$

where $n$ is the number of similar assets found in the historical dataset; $R_i$ is the average aging rate of asset $i$ observed at the initial time point; $R_i^T$ is the average aging rate of asset $i$ observed after time $T$; $R$ is the current aging rate of target asset. Following (15), the future conditional age of the target asset can be estimated by:

$$A^C = R^T \times (A^P + T) \quad (17)$$

It should be noted that sometimes the long-term asset condition dataset may not have records for the desired time interval. For example, distribution poles have been historically inspected every 10 years but the task now is to predict a pole's status in 5 years. In cases like this, linear interpolation can be used to derive the aging rate in 5 years. Mathematically, it is given as:

$$R_i^{T'} = R_i + \frac{(R_i^T - R_i)T'}{T} \quad (18)$$

where $R_i$ is the average aging rate of asset $i$ observed at the initial time point; $R_i^T$ is the historical aging rate of asset $i$ observed after time $T$; $T'$ is desired time increment for prediction. After this step, $R_i^{T'}$ can be taken into (16) to replace $R_i^T$. Similarly, $A^C$ can be calculated by using (17).

The above method is very powerful because it traces historical aging rate variation of similar assets for the prediction of the target asset in the future.

### C. Predict asset future status and failure probability

After the previous steps, both future physical age and conditional age of the target asset are obtained. The two input values can now be fed into the LRC which is previously trained in the learning process. In return, LRC will automatically classify this asset to either working or failed status. To this point, individual asset status prediction is completed.

In addition to predicting asset status as either working or failed, LRC can also be used to calculate the failure probability of asset using equation (13). On top of the calculated failure probabilities, Monte Carlo simulation and risk analysis that take asset failure consequence data into consideration can be performed too, similar to the risk analysis that uses Weibull distribution functions [1],[16].

### V. CASE STUDIES

The proposed approach was applied to an urban distribution system in Western Canada. 1000 single-phase 13.8kV XLPE cable segments recorded 5-year apart (in 2012 and 2017) were included in the case studies. Three condition features were gathered by both proactive and reactive tests and are explained as below:

- Partial discharge (PD) test result: Voids and trees in the insulation, moisture filtration and other hazardous conditions can lead to PD activities inside cable [31]. Having excessive PD is an early sign of cable failures. In this utility company, PD severity is measured periodically and reactively after a cable fault. Measured results were converted to a numerical condition feature.
- Neutral corrosion test result: neutral corrosion condition is tested using time domain reflectometer (TDR) periodically and reactively after a cable fault [32]. Corroded neutral can also lead to a cable failure. This is also a numerical health condition feature.
- Visual inspection result: This manual inspection looks for cable discoloration, surface cracks, and surface contamination. Tactile information on surface texture and rigidity is also considered [33]. In the end, a health rating in the choices of poor, medium and good is given. This is a categorical condition feature.

### A. Classification evaluation

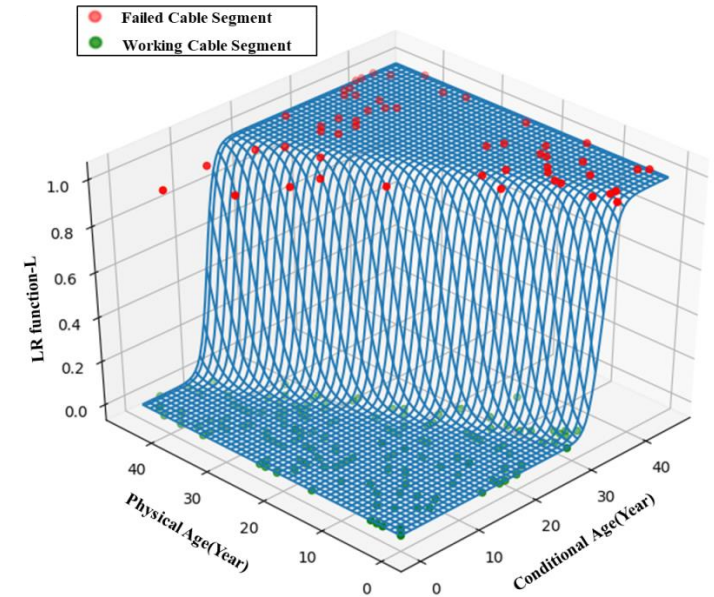

Fig.3. The trained logistic regression classifier

The first evaluation is on the model's classification performance. Following the learning process discussed in Section III, the 2012 data was split into training set and test set based on an 80:20 ratio. The test set was used to evaluate the model performance. The trained LRC is plotted in Fig.3: The meshed surface indicates the logistic regression value for any pair of physical and conditional ages. Red and green points represent the actual failed and working cable segments in the test set.

Confusion matrix, precision index, recall index and F1-Score were used for evaluation. In modern data science, these are widely applied evaluation tools for binary classification and prediction tasks [34]. These indices are able to describe how well the model classifies or predicts two statuses separately.

The confusion matrix is shown in Table IV. TP, FP, TN and FN stand for true positive, false positive, true negative and false negative counts.

TABLE IV: CONFUSION MATRIX FOR CLASSIFICATION EVALUATION

| Total N=200 | Predicted: Positive(Failed) | Predicted: Negative(Working) |
|---|---|---|
| Actual: Positive(Failed) | TP=45 | FN=6 |
| Actual: Negative(Working) | FP=5 | TN=144 |

For asset failed status, the precision index and the recall index are:

$$\begin{cases} Precision = \frac{TP}{TP+FP} \\ Recall = \frac{TP}{TP+FN} \end{cases} \quad (19)$$

For asset working status, the precision index and the recall index are:

$$\begin{cases} Precision = \frac{TN}{TN+FN} \\ Recall = \frac{TN}{TN+FP} \end{cases} \quad (20)$$

The F1-Score is the harmonic mean of Precision and Recall:

$$F1 = 2\frac{Presision \cdot Recall}{Presision+Recall} \quad (21)$$

Table V summarizes the precision, recall and F1-Score results following (19)-(21).

TABLE V: PRECISION, RECALL AND F1-SCORE FOR CLASSIFICATION EVALUATION

| Evaluation Category | Precision | Recall | F1-Score |
|---|---|---|---|
| Asset Failed Status | 0.90 | 0.88 | 0.89 |
| Asset Working Status | 0.96 | 0.97 | 0.96 |
| Average | 0.93 | 0.92 | 0.93 |

As can be seen in Table V, the trained LRC is an excellent classifier for classifying asset statuses using conditional age and physical age. The performance for classifying failed assets is slightly lower than classifying working assets. Also, the classifier is slightly more conservative when classifying an asset as failed because the recall 0.88 is lower than the precision 0.90.

*B. Prediction evaluation using one-time asset condition data*

Here, it is assumed that only 2012 dataset is known before prediction. (14) and (15) were used for estimating future conditional age. Again, 20% of the data was used for performance evaluation. The predicted asset operation statuses were compared with the actual 2017 asset operation statuses. The confusion matrix and evaluation indexes are summarized in Table VI and Table VII. Compared to the classification results in Table IV and Table V, the prediction performance values are lower. This is expected because prediction is to classify assets in the future when the future conditional ages are unknown and have to be estimated. Still, the average precision, recall and F1-score are satisfactory, even when using one-time asset condition data.

TABLE VI: CONFUSION MATRIX FOR PREDICTION EVALUATION (ONE-TIME ASSET CONDITION DATA)

| Total N=200 | Predicted: Positive(Failed) | Predicted: Negative(Working) |
|---|---|---|
| Actual: Positive(Failed) | TP=41 | FN=15 |
| Actual: Negative(Working) | FP=11 | TN=133 |

TABLE VII: PRECISION, RECALL AND F1-SCORE FOR PREDICTION EVALUATION (ONE-TIME ASSET CONDITION DATA)

| Evaluation Category | Precision | Recall | F1-Score |
|---|---|---|---|
| Asset Failed Status | 0.79 | 0.73 | 0.76 |
| Asset Working Status | 0.90 | 0.92 | 0.91 |
| Average | 0.84 | 0.83 | 0.84 |

*C. Prediction evaluation using long-term asset condition data*

In this case, assuming both 2012 and 2017 records were known, the more accurate conditional age estimation method with (16) and (17) was used. As shown in Table VIII and IX, the prediction performance is better than Section V-B.

TABLE VIII: CONFUSION MATRIX FOF PREDICTION EVALUATION (LONG-TERM ASSET CONDITION DATA)

| Total N=200 | Predicted: Positive(Failed) | Predicted: Negative(Working) |
|---|---|---|
| Actual: Positive(Failed) | TP=47 | FN=9 |
| Actual: Negative(Working) | FP=9 | TN=135 |

TABLE IX: PRECISION, RECALL AND F1-SCORE FOR PREDICTION EVALUATION (LONG-TERM ASSET CONDITION DATA)

| Evaluation Category | Precision | Recall | F1-Score |
|---|---|---|---|
| Asset Failed Status | 0.84 | 0.84 | 0.84 |
| Asset Working Status | 0.94 | 0.94 | 0.94 |
| Average | 0.89 | 0.89 | 0.89 |

*D. Prediction evaluation using Weibull Distribution*

Standard Weibull distribution based prediction is used for comparison. In the 2012 dataset, 80% of physical age and asset status data was used to develop the Weibull distribution function. Then 20% of the data was used for prediction in 2017 and compared to the actual 2017 asset statuses. Similar to LRC, if the failure probability is greater than 0.5, it is classified as a failed asset otherwise a working asset.

As shown in the Table X and XI, the average precision, recall and F1-Score are all lower than the previous results of LRC which used both physical age and condition data. Furthermore, it is observed that the sum of TP and FP are much higher than in Table VI and Table VIII; also, the precision 0.64 is quite low. This means Weibull distribution in this application example are more inclined to predict an asset as a failed asset. In comparison, the proposed LRC is less biased and more accurate. In the case of using one-time asset condition data, the average F1-Score increased from 0.78 to 0.84; in the case of using long-term asset condition data, it increased from 0.78 to 0.89. The better performance of LRC is due to the utilization of asset condition data in addition to just using physical age. As a result, the prediction becomes more specific and accurate for individual assets.

TABLE X: CONFUSION MATRIX FOR PREDICTION EVALUATION (WEIBULL DISTRIBUTION)

| Total N=200 | Predicted: Positive(Failed) | Predicted: Negative(Working) |
|---|---|---|
| Actual: Positive(Failed) | TP=44 | FN=12 |
| Actual: Negative(Working) | FP=25 | TN=119 |

TABLE XI: PRECISION, RECALL AND F1-SCORE FOR PREDICTION EVALUATION (WEIBULL DISTRIBUTION)

| Evaluation Category | Precision | Recall | F1-Score |
|---|---|---|---|
| Asset Failed Status | 0.64 | 0.79 | 0.70 |
| Asset Working Status | 0.91 | 0.83 | 0.87 |
| Average | 0.77 | 0.81 | 0.78 |

*E. Sensitivity of data quality*

Since the proposed approach is a data-driven approach, we also tested the sensitivity of data quality. 4 types of data noises are purposely introduced to the original 1000-record dataset in the following ways: the asset operation statuses of 5 randomly selected records are swapped between "Working" and "Failed"; the asset operation statuses of 10 randomly selected records are swapped; the values of 5 randomly selected numerical health condition features are increased by 50%; the values of 10 randomly selected numerical health condition features are increased by 50%.

Following the same long-term prediction process discussed above, Table XII shows that the prediction accuracy decreases after the data quality is artificially reduced. The accuracy is more susceptible to the distortion caused by swapping asset status.

TABLE XII: SENSITIVITY OF DATA QAULITY

| Data Quality | Original Dataset | Swapping 5 Asset Statuses | Swapping 10 Asset Statuses | Increasing 5 Feature Values | Increasing 10 Feature Values |
|---|---|---|---|---|---|
| Average F1-Score | 0.89 | 0.83 | 0.75 | 0.87 | 0.83 |

*F. Sensitivity of data size*

The impact of data size on the prediction performance was also tested. The average F1-Scores resulted from using the original complete 1000 records, a randomly selected subset of 750 records, a subset of 500 records and a subset of 250 records are listed in Table XIII.

TABLE XIII: SENSITIVITY OF DATA SIZE

| Data size | 1000 | 750 | 500 | 250 |
|---|---|---|---|---|
| Average F1-Score | 0.89 | 0.84 | 0.76 | 0.66 |

As expected, the average F1-Score decreases as the data size decreases.

## VI. DISCUSSION AND CONCLUSION

This paper presents a novel and comprehensive data-driven approach for predicting asset failures based on asset condition data. As more and more utility companies are turning to condition based maintenance, it is important to research how to incorporate asset condition data into asset class failure prediction, risk analysis and asset strategy optimization. This paper made an attempt in this direction. The main contributions of the proposed approach are:

- It includes a unique learning process to analyze the inner structure of historical asset condition data and establish the statistical relationship between an asset's operation status and its conditions in conjunction with physical age;
- It includes a unique prediction process to predict the change of asset conditions over time and future asset status based on newly proposed concepts asset conditional age and average aging rate;
- It uses sophisticated unsupervised and supervised learning methods instead of Weibull distribution for asset class failure prediction;
- It can effectively incorporate the asset condition data to produce more accurate and specific prediction for individual assets in an asset class.

Detailed case studies are given and this method demonstrates superior performance measured by F1-Score over traditional Weibull distribution based method. It is also a practical approach that utility engineers can apply in real world.

Since the proposed method is a data-driven method, the results by nature will be sensitive to data quality and data size, as manifested in Section V-E and Section V-F. In practice, there are generally two ways to ensure data quality for the discussed application: proactive filtering and reactive filtering. When an inspection record is produced and ready to be logged into the CMMS database, it is recommended that maintenance engineers check the data quality immediately. This includes checking data errors such as zeros, values that exceed the empirical ranges or values that contradict with asset domain knowledge. If a problem is identified, instead of logging this inspection record into the database, the record needs to be either corrected or flagged as flawed. This practice will ensure the data quality at the data-entry stage. This method is called proactive filtering; the other method is reactive filtering. This is the process of retroactively checking the data quality and identifying data outliers. This method is not as timely as proactive filtering. But it has its advantage since it can rely on statistical information such as max/min/average of previously gathered condition values or even probabilistic distributions of these values to identify outliers. It should be noted that the proposed learning process does not have to use all historical data points but rather statistically representative data points – this suggests that outliers that fall out of the typical ranges can be excluded from the learning process. After all, the goal is to produce a statistically confident prediction rather than the prediction of a few outliers. Through proactive and reactive filtering, data outliers can be effectively identified and managed for the proposed work flow. For data size, Section V-F reveals a fact that the prediction accuracy will decrease when the number of points used for the learning process decreases. It should be noted that this problem is also true for Weibull distribution based methods. On the bright side, one general trend of utility companies is the increasing adoption of sophisticated CMMS system to collect, store and utilize data for asset management. As we progress, more and more asset data will become available to support this kind of application.

Another exciting development in the big data and artificial intelligence field is Natural Language Processing and Semantic Analysis [35]. This type of development may help utility companies to analyze irregular text based inspection records produced by field crews and convert them to quantitative data that can be directly utilized by the proposed approach.

This paper focuses on establishing the theoretical framework and proving the feasibility and benefits of the propose approach. In the future, when more asset condition data becomes available, more asset classes will be tested using the proposed approach. Other types of classifiers can be evaluated in comparison with LRC. Furthermore, other clustering methods such as DBSCAN and Gaussian Mixture Models can also be investigated [36-37].

**Ming Dong** (S'08, M'13, SM'18) received his Ph.D degree from Department of Electrical and Computer Engineering, University of Alberta, Canada in 2013. Since graduation, he has been working in various roles in two major electric utility companies in West Canada as a registered Professional Engineer (P.Eng.) and Senior R&D Engineer for 6 years. Dr. Dong was the recipient of the Certificate of Data Science and Big Data Analytics from Massachusetts Institute of Technology. He is also a regional officer of Alberta Artificial Intelligence Association. His research interests include applications of artificial intelligence and big data technologies in power system planning and operation, power quality data analytics and non-intrusive load monitoring.